%Paper: astro-ph/9305037
%From: ld@cp.dias.ie
%Date: Fri, 28 May 93 15:30:52 BST

%
% Uses the Astronomy and Astrophysics style macros (plain TeX version)
% e-mail requests/comments to ld@cp.dias.ie
%
\input aa.cmm
\MAINTITLE{The gamma-ray visibility of supernova remnants}
\SUBTITLE{A test of cosmic ray origin}
\AUTHOR{L. O'C. Drury@1, F. A. Aharonian@2 and H. J. V\"olk@2}
\INSTITUTE{
@1
Dublin Institute for Advanced Studies,
School of Cosmic Physics,
5 Merrion Square,
Dublin 2,
Ireland
@2
Max-Planck-Institut f\"ur Kernphysik,
Postfach 103980,
W-6900 Heidelberg-1
}

\OFFPRINTS {L. O'C. Drury}
\DATE{Received ........; accepted ..........}
\ABSTRACT{
Recent calculations of particle acceleration in supernova remnants
(SNRs) are used to estimate the associated $\gamma$-ray production.
For source spectra which are power-laws in momentum (or rigidity)
the production efficiency of $\gamma$-rays with
energy $E_\gamma>100\,\rm MeV$ is shown to be about a factor 2--3
lower than the value conventionally used for the interstellar medium
and to depend only weakly on the spectral index of the power-law
(in the range expected). Because the energy transferred to accelerated
particles is rather tightly constrained by the total Galactic cosmic ray
power, if SNRs are the main source of Galactic cosmic rays, this leads
to an almost model-independent prediction of the SNR $\gamma$-ray luminosity
in the band $E_\gamma>100\,\rm MeV$.

A detailed discussion of instrumental sensitivities and backgrounds shows
that detection of SNRs in the $E_\gamma>100\,\rm MeV$ band with, for
example, the Energetic Gamma Ray Experiment Telescope (EGRET)
will be difficult, but should not be
impossible. However, and significantly, the prospects look
much better in the TeV band accessible to modern imaging atmospheric
Cherenkov telescopes. It should be possible to detect SNRs out to
distances of $10\,\rm kpc$ if the region of the ISM into which they
are expanding has a high enough density ($n>0.1\rm\,cm^{-3}$) so that
their $\gamma$-ray luminosity is high enough.

Finally, it is pointed out that existing and planned air-shower arrays
can place important limits on the extension of the accelerated
particle spectra in SNRs to energies above $100\,\rm TeV$. In conjunction
with spectral measurements in the TeV region and detections or upper
limits in the $100\,\rm MeV$ band this could provide a
crucial test of current theories of particle acceleration in SNRs.}
\THESAURUS{1.01.1, 3.10.1, 7.36.1, 19.92.1}

\maketitle

\titlea{Introduction}

There is no clear observational proof that the nuclear component of
cosmic rays is produced in supernova remnants (SNRs) although this is
widely believed, at least for particle energies less than about
$10^{14}\,\rm eV$ per nucleon. This general conviction is mainly
based on two
facts; first, that a theoretical acceleration mechanism exists with
certain attractive properties (diffusive shock acceleration:
two recent reviews which emphasise the astrophysical applications
are Blandford and Eichler, 1987; Berezhko and Krymsky, 1988),
and secondly, that the galactic SNae (and thus the resultant SNRs) are
almost the only known potential sources with the necessary amount of
available energy. Although plausible, these are indirect and
unsatisfactory arguments. It is clearly desirable to seek direct
evidence for accelerated nuclei in SNRs.
The best hope would appear to be gamma-ray observations.
Indeed some thirty years ago it was this, in the context of cosmic ray
sources in general rather than SNRs in particular,
which was at the heart of the rationale for missions like SAS-II and COS-B.
If the nuclear
component of the cosmic rays is strongly enhanced inside SNRs,
then through nuclear collisions leading to pion production and
subsequent decay, $\gamma$-rays will inevitably be produced\fonote{There
will also be Bremsstrahlung and Inverse Compton components,
produced by the electrons, both primary (directly accelerated)
and secondary (from $\pi^\pm$ decays), known from radio synchrotron
observations to be present in SNRs. By considering only the gamma
rays of hadronic origin we obtain a lower bound to the overall
$\gamma$-ray emission of SNRs.}.
In this paper we discuss some estimates of the
$\gamma$-ray visibility of SNRs in the energy range around $100\,\rm MeV$
typical of the EGRET instrument on the Compton Observatory, the TeV
energy range accessible to ground-based Cherenkov telescopes and the
range above $10\,\rm TeV$ which can be covered by modern airshower arrays.

Diffusive shock acceleration, as applied to cosmic ray production in
SNRs has been studied by many authors
(eg Blandford and Ostriker, 1980;
Krymsky and Petukhov, 1981;
Prischep and Ptuskin, 1981;
Bogdan and V\"olk, 1983;
Moraal and Axford, 1983;
Lagage and Cesarsky, 1983;
Jokipii and Ko, 1987;
Berezhko and Krymsky, 1988;
V\"olk, Zank and Zank, 1988;
Dorfi, 1990;
Kang and Jones, 1991). In general, it must be
said that all calculations contain {\it ad hoc} parametrizations of
various aspects of the physics, especially the injection of particles
into the acceleration process. Recent comparisons (Kang and Drury, 1992)
show that the
`simplified models' (Drury, Markiewicz and V\"olk, 1989 [DMV]; Markiewicz,
Drury and V\"olk 1990) are good approximations to
the much more detailed hydrodynamical calculations, and in particular
that the differences (in the worst cases factors of order two in the
energy densities)
are less than the uncertainties in the
parametrizations of the underlying
physics. This suggests that they can be used
to obtain estimates of the gamma-ray flux to be expected
from nuclear interactions in a SNR.

It is important to note that although the injection rate  has to
be specified in an essentially arbitrary fashion, the overall rate is rather
tightly constrained by the observationally inferred energy requirements.
When allowance is made for the energy-dependent escape of particles above
a few GeV (Swordy et al, 1990) the traditional estimates of production
efficiency are raised from a few percent to something above 10\% per
supernova (V\"olk et al, 1985; DMV) if the bulk of the cosmic ray
nuclei are indeed accelerated in SNRs.

\titlea{Expected $\gamma$-ray fluxes}

In the simplified models the SNR is divided into
three regions:  an interior region filled with hot gas and accelerated
particles but very little mass, an immediate post shock region where
most of the matter is concentrated, and a shock precursor region where
the accelerated particles diffusing ahead of the shock
compress the ambient medium.  In each of these
regions we know the energy density of the accelerated cosmic ray
particles and the mass density.
The production rate of $\gamma$-rays per unit volume can be written
as
$$
Q_\gamma = {\cal E}_\gamma n = q_\gamma n E_C
\eqno\autnum
$$
where $n$ is the number density of the gas, $E_C$ is the CR energy
density and $q_\gamma$ is the $\gamma$-ray emissivity normalized to the
CR energy density, $q_\gamma={\cal E}_\gamma/E_C$.
For the galactic CR in the diffuse interstellar medium
(ISM), outside sources, the $\gamma$-ray emissivity above $100\,\rm MeV$ due
to nuclear collisions is about ${\cal E}_\gamma(\ge 100{\,\rm MeV})
\approx (1 - 1.5)\times10^{-25}\,\rm s^{-1}(\hbox{H-atom})^{-1}$
(e.g.~Stecker, 1971; Dermer, 1986) and $E_C\approx 10^{-12}\,\rm erg\,cm^{-3}$
The corresponding $q_\gamma(\ge 100\,\rm MeV)\approx
10^{-13}\,\rm s^{-1}erg^{-1}cm^3(\hbox{H-atom})^{-1}$ in the ISM can also be
used for rough estimates of the $\gamma$-ray luminosity of SNRs (see below).

In fact, because the spectrum in the source is expected to be harder
than the observed cosmic ray spectrum (which we know, from the energy
dependence of the relative abundance of spallation products,
to have been softened by propagation effects;
Juliusson et al, 1972) this is almost certainly an
overestimate of the production rate in SNRs. The
spatially averaged spectrum in the source is
probably of the form $f(p)\propto p^{-\alpha}$ with $4\le\alpha\le4.3$,
whereas the observed cosmic ray spectrum has $\alpha\approx 4.7$.
Gamma rays above $100\,\rm MeV$
are produced by particles of energy above about $1\,\rm GeV$ per nucleon
and the cross-sections are not very energy-dependent; thus the gamma production
rate may be taken to be proportional to the number density of relativistic
particles. For a given energy density, the harder the spectrum the lower
will be the total number density of particles. On the other hand, if the
spectrum extends to energies below about $1\,\rm GeV$ per nucleon, a
harder spectrum will have fewer particles which contribute to the total
energy density but do not generate observable  $\gamma$-rays.
To quantify these competing effects let us,
following DMV, consider a simple power-law spectrum
of the form
$$
f(p)=f_0(p/2mc)^{-\alpha}
\eqno\autnum
$$
where $f$ is the (isotropic) phase space density and $p$ is momentum
with a lower cut-off at $p_{\rm
inj}$ and an upper cut-off at $p_{\rm max}$. Then the number density
of relativistic particles with $p>2mc$ is easily calculated to be
$$
N_{R}={1\over\alpha-3} 4\pi f_0 (2mc)^3
\left[1- \left(p_{\rm max}\over m c\right)^{3-\alpha}\right]
\eqno\autnum
$$
and for $\alpha \ge 4$ and $p_{\rm max}\gg mc$ the dependence on the upper
cut-off can be neglected.
On dividing by the corresponding energy density (as calculated in DMV)
we obtain
$$
\eqalign{
{N_R\over E_C} &=
{1\over 2mc^2}\left[1-{p_{\rm inj}\over 2mc} + \ln\left(p_{\rm max}\over
2 mc\right)\right]^{-1},\qquad \alpha=4,\cr
&=
{1\over2(\alpha - 3)mc^2}
\left[{1-(p_{\rm inj}/2mc)^{5-\alpha}\over 5 - \alpha}
\right.\cr
&\qquad\left. +{(p_{\rm max}/2mc)^{4-\alpha}-1\over 4 - \alpha}\right]^{-1}
,\qquad 4<\alpha<5,\cr
&={1\over 4 mc^2}\left[\ln\left(2mc\over p_{\rm inj}\right) + 1 -
{2mc\over p_{\rm max}}\right]^{-1},\qquad \alpha=5.\cr
}
\eqno\autnum
$$

It is easy to see that for typical values of $p_{\rm inj}/mc
\approx 10^{-3} - 10^{-2}$ and $p_{\rm max}/mc \approx 10^5 - 10^6$
the ratio $N_R/E_C$ is rather independent of $\alpha$ even in the
most extreme case of a hard source spectrum proportional to $p^{-4}$
and an ambient spectrum of the form $p^{-4.7}$.
This semi-quantitative argument is confirmed by numerical calculations
which show that for the spectrum of accelerated particles
represented by Eq.~(2) the
production rate is almost independent of the spectral index $\alpha$.
In fact, for $4.1\le \alpha \le 4.7$ the production rate may be
considered constant within 20\% (see Table 1) of
$$
q_\gamma(\ge 100\,{\rm MeV})\approx 0.5\times 10^{-13}
\rm\,s^{-1}erg^{-1}cm^3(\hbox{H-atom})^{-1}
\eqno\autnum
$$
which is a factor of three less than the corresponding value for a
Galactic CR flux of the form
$J_p(E)\propto (E+mc^2)^{-2.75}$ where $E$ is kinetic energy.
This particular form for the spectrum gives a very high (close indeed to
the maximum possible) efficiency for the production of $\ge100\rm\,MeV$
$\gamma$-rays (at fixed CR energy density) since it contains negligible
energy density in nonrelativistic particles (useless for gamma production)
and, as mentioned above, at the same time has a low energy density
for particles with energy much above a few GeV. As the $\pi^0$-production
cross-section becomes constant above a few GeV the maximum value is
reached in the case of a monoenergetic beam of protons with energy
1--$2\,\rm GeV$. The results of numerical calculations of
$q_\gamma(\ge 100\,{\rm MeV})$ for different spectral indices $\alpha$
are presented in Table 1.

At energies $E\gg 100\,\rm MeV$ the production spectrum of $\gamma$-rays
reproduces the spectral shape of the distribution of parent protons.
The emissivity, ${\cal E}_\gamma$, for power-law spectra of relativistic
protons can be presented in a simple analytic form (Appendix A).
In the case of a source spectrum of the form given in Eq.~(2), the
production rate, normalized to the energy density of the accelerated
particles is
$$
q_\gamma(>E)=q_\gamma(\ge 1{\,\rm TeV})
\left(E\over 1{\,\rm TeV}\right)^{3-\alpha}
\eqno\autnum
$$
where $q_\gamma(\ge 1{\,\rm TeV})$ is also tabulated in Table 1 for
various spectral indices $\alpha$.

\begtabfull
\tabcap{1}{The production rates of $\gamma$-rays
for different spectral
indices $\alpha$ of the parent cosmic ray distributions,
$f(p)\propto p^{-\alpha}$. The contribution of nuclei other than
H in both the CR and the target matter
is assumed to be the same as in the ISM, ie the pure proton
contribution is multiplied by a factor 1.5 (cf Dermer, 1986).
In calculating the low energy part of the production rate we
used the formalism developed by Stecker (1971).
The units are $\rm s^{-1} erg^{-1} cm^3 (\hbox{H-atom})^{-1}$.}
\halign{#\hfil&&\quad#\hfil\cr
\noalign{\hrule\medskip}
$\alpha$ &
$q_\gamma(\ge 100{\,\rm MeV})$ &
$q_\gamma(\ge 1{\,\rm TeV})$ \cr
\noalign{\medskip\hrule\medskip}
4.1 & $0.46\times 10^{-13}$ & $1.02\times 10^{-17}$ \cr
4.2 & $0.58\times 10^{-13}$ & $4.9 \times 10^{-18}$ \cr
4.3 & $0.61\times 10^{-13}$ & $2.1 \times 10^{-18}$ \cr
4.4 & $0.57\times 10^{-13}$ & $8.1 \times 10^{-19}$ \cr
4.5 & $0.51\times 10^{-13}$ & $3.0 \times 10^{-19}$ \cr
4.6 & $0.44\times 10^{-13}$ & $1.0 \times 10^{-19}$ \cr
4.7 & $0.39\times 10^{-13}$ & $3.7 \times 10^{-20}$ \cr
\noalign{\medskip\hrule}}\endtab

The simplified models do not as yet include any cooling of the gas
(in contrast to the hydrodynamic models of Dorfi, 1991a,b), however we do
not believe that this is important for the estimates presented here.
The cosmic ray energy density, because of efficient diffusion, is rather
uniformly distributed inside the remnants so that clumping of the gas as
a result of cooling is not likely significantly to change the results.
Of course cooling does change the shock strength, and therefore the
acceleration efficiency, in the late stages of SNR evolution.

\begfigwid 18 cm
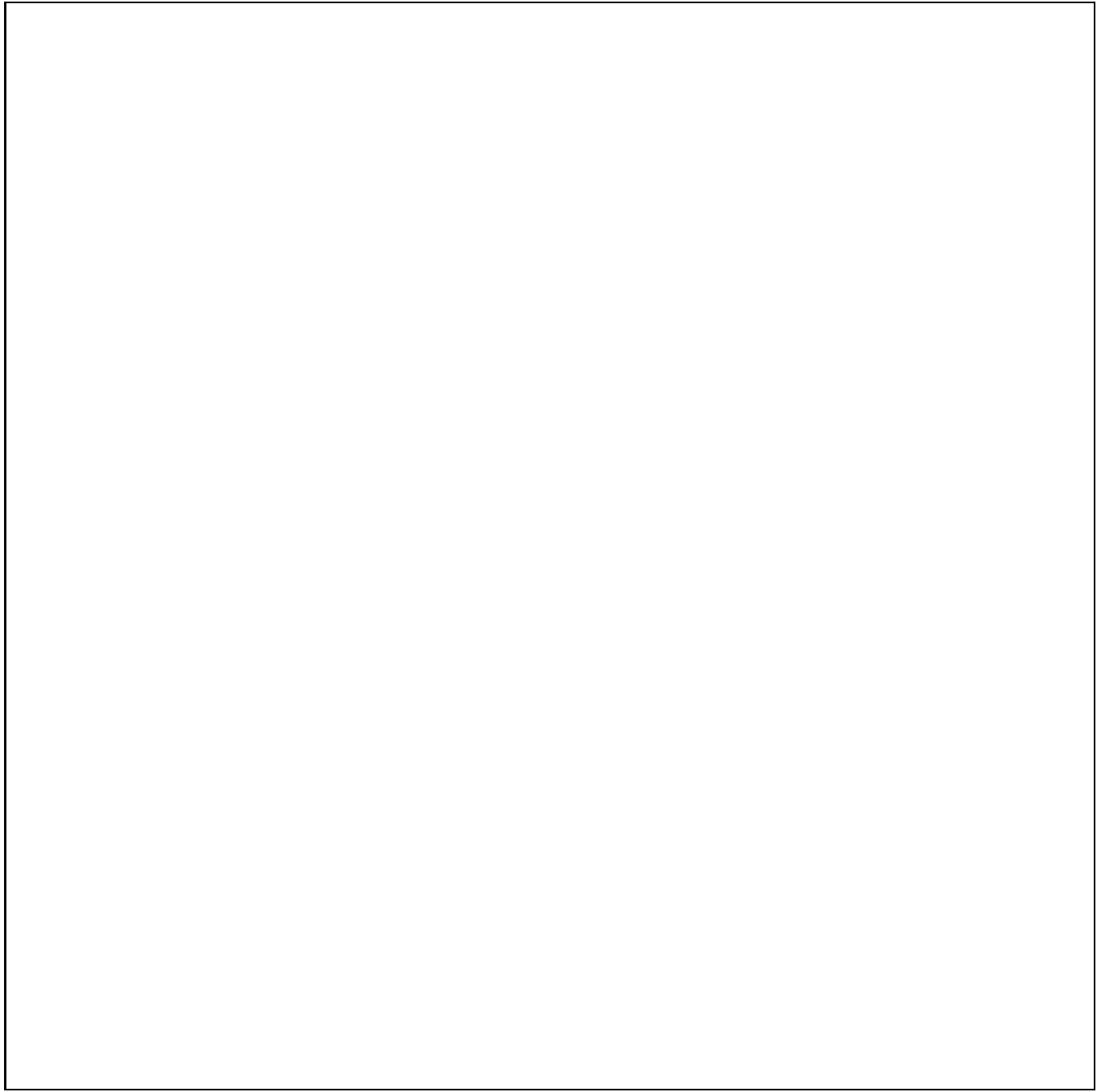
\figure{1}{The evolution, as a function of age in years, of the total
$\gamma$-ray luminosity (in photons with energy above $100\,\rm MeV$
per second) and
the central surface brightness (in photons $\rm cm^{-2}sr^{-1}s^{-1}$)
together with the shock radius and the width of the precursor region
(in pc) and the partition of the explosion energy between accelerated
particles ($E_C$) and thermal gas ($E_G$) for a SNR expanding into
a medium of density $0.1\,\rm cm^{-3}$.}
\endfig

To discuss the visibility of SNR as $\gamma$-ray sources two quantities
seem most interesting. The first is the total luminosity of the remnant
in photons per second, the second the surface brightness along a line
through the centre of the remnant in photons $\rm
cm^{-2}\,s^{-1}\,sr^{-1}$. While the first measure is appropriate to
discussions of the detectability of unresolved sources, the second is
more appropriate for spatially resolved sources. Figure~1 shows both
measures for a rather standard SNR model together with the radii
of the shock, the thickness of the precursor and the partition of
the explosion energy between thermal and nonthermal particles.
The dotted vertical line indicates the point where the post-shock
gas temperature drops below $10^6\,\rm K$ and cooling becomes
potentially important (although as noted we have not included
cooling in the present calculations).  For this calculation the ejecta
mass was set to $1\,\rm M_\odot$ and the ambient density to $0.1 \,\rm
cm^{-3}$. Alfv\'en heating in the precursor and diffusive coupling
between the interior of the remnant and the precursor were included
with parameters $\alpha_H=1$ and $\alpha_C=10$; the
injection was parameterized by specifying that a constant fraction
$\epsilon=10^{-3}$ of the kinetic energy dissipated in the shock was
given to energetic particles capable of further acceleration.
The physical processes represented by the parameters $\epsilon$,
$\alpha_H$ and $\alpha_C$ are discussed in detail in DMV. Here
we note simply that the effect quantified by $\alpha_C$ is
clearly required on theoretical grounds and has been shown to
be a significant effect in recent numerical studies (eg Dorfi, 1991a,b). The
importance of Alfv\'enic effects is the subject of a recent study by
Jones (1993). It is worth noting that both effects act to reduce the
amount of acceleration and thus tend to decrease the $\gamma$-ray
production.

The total gamma-ray luminosity is given by $\int q_{\gamma} n E_C\,d^3r$
which, in the context of the simplified models, can be written as
$q_\gamma\left(M_1E_{C1}+M_2E_{C2}\right)$ where $M_1$ is the total
mass in the precursor region, $M_2$ that in the immediate post-shock
region and $E_{C1,2}$ are the corresponding cosmic ray energy densities.
Because of diffusion across the shock the two energy densities are in fact
equal, $E_{C1}=E_{C2}$. The common value is also not very different
from $E_{C3}$, the energy density in the interior of the remnant, for
two reasons. First, there is diffusive coupling between the acceleration
region around the shock and the interior of the remnant. Secondly, if
the acceleration is efficient, the cosmic rays provide a substantial, if
not the dominant, part of the interior pressure and the interior of the
remnant has, for dynamical reasons, to be in rough pressure equilibrium.
It follows that, to order of magnitude, the cosmic ray energy density
throughout the remnant and in the shock precursor is
$E_{C1}=E_{C2}\approx E_{C3}\approx 3 \theta E_{SN}/4\pi R^3$ where $\theta$ is
the fraction of the total supernova explosion energy, $E_{\rm SN}$,
converted to cosmic ray energy and $R$ is the remnant radius.
Thus for the $\gamma$-ray luminosity we
can estimate
$$
\eqalign{
L_\gamma &= q_\gamma (M_1+M_2) {3 \theta E_{\rm SN}\over 4\pi R^3}
\approx \theta q_\gamma E_{\rm SN} n\cr
&\approx  10^{38} \theta (E_{\rm SN}/10^{51}{\,\rm erg})
(n/1\,\rm cm^{-3}) \, ph\,s^{-1}\cr}
\eqno\autnum
$$
where $n$ is the ambient density.
Obviously the exact value of $\theta$ depends on the details of the
model, in particular the injection parametrization used, but clearly
$\theta <1$ and, as discussed above, the observations appear to require
$\theta \ge 0.1 $ at the point when acceleration stops and the produced
cosmic rays diffuse out into the ISM.
In fact for reasonably plausible injection models
$\theta$ is roughly constant throughout the Sedov phase with only
a moderate dependence on external parameters such as the ambient density
(Markiewicz et al, 1990). At earlier times, during the free-expansion
phase, the amount of energy processed through shocks is small (almost
all of the explosion energy is in the form of kinetic energy of the
ejecta) and $\theta \ll 1$.

The calculated results agree well
with this simple picture; the total luminosity is low during the
free-expansion phase, but is roughly constant during the Sedov phase.
The Sedov phase (or better, Sedov-like, because the models are not
exact similarity solutions) starts when the amount of swept up
matter is roughly equal to the ejecta mass. Typically this occurs
at a radius $O(10{\,\rm pc})$ and the SNR spends most of its useful
life (until the outer shock becomes weak) in this phase.
Cooling of the post-shock gas is important when the post-shock
temperature drops below about $10^6\,\rm K$, but is not as
catastrophic for these models as for pure gas models because the
CR do not cool and continue to provide an internal pressure in
the remnant. As long as the outer bounding shock is strong
and active in accelerating particles the diffusion
coefficient for cosmic rays near the shock front is very small
($\ll R \dot R$ except at the highest energies). Thus the
accelerated particles are confined to the remnant and only
diffuse out into the ISM when the shock, either because of
expansion or cooling, is no longer able to maintain the strong
levels of magnetic disturbance which lead to diffusion
coefficients very much smaller than those in the ambient ISM.

\begfigwid 18 cm
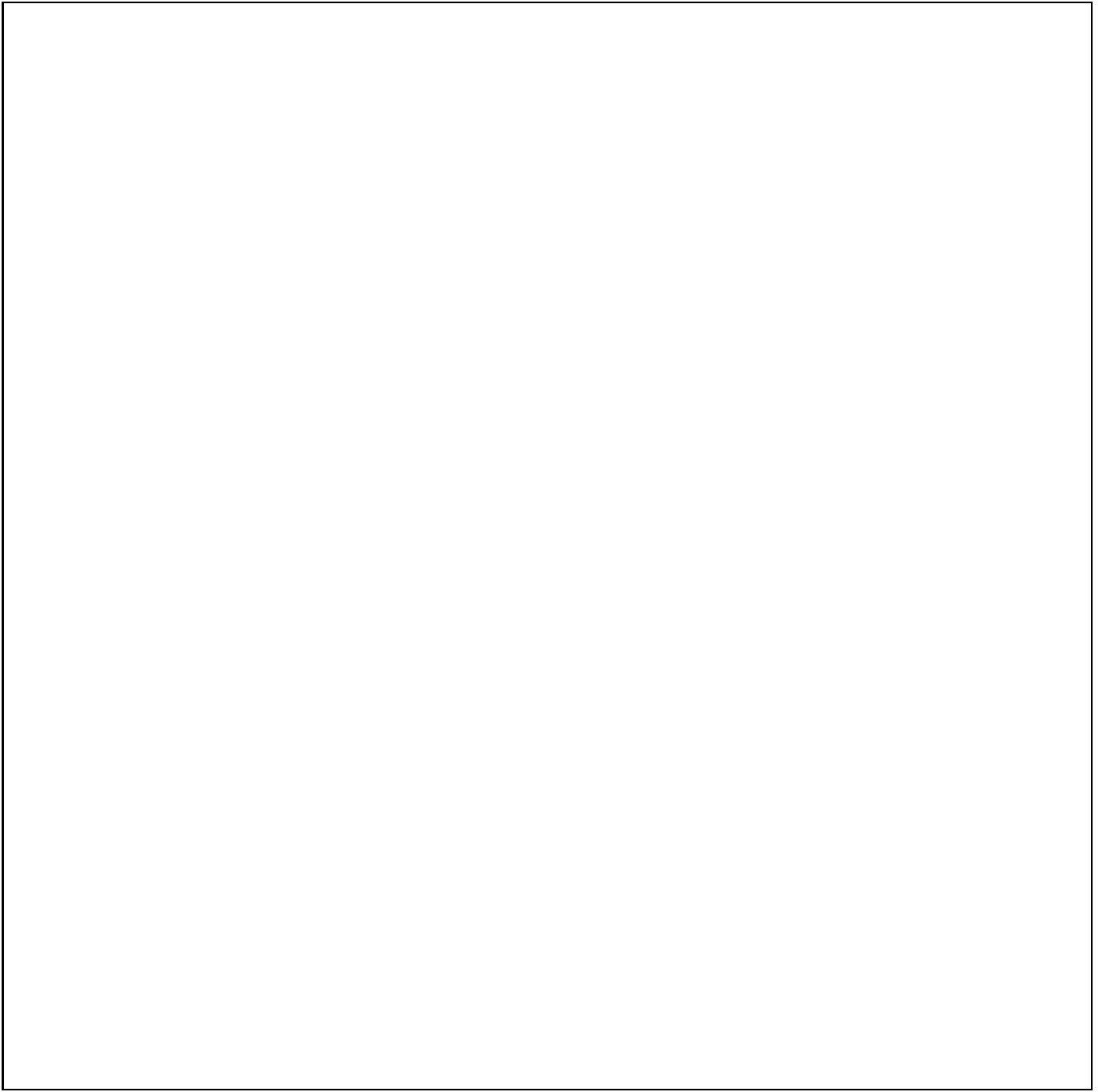
\figure{2} {As Fig.~1, but for an external density of $n=5\,\rm cm^{-3}$.}
\endfig

If the SNR is a distance $d$ away, this translates into a flux at
the earth of
$$
\eqalign{
F(\ge100{\,\rm MeV})\approx  4.4\times 10^{-7} \theta &
\left(E_{\rm SN}\over 10^{51}\rm \,erg\right)
\left(d\over 1\,\rm kpc\right)^{-2} \cr
& \left(n\over\rm1\,cm^{-3}\right)\rm\, cm^{-2}s^{-1}\cr}
\eqno\autnum
$$
$\gamma$-rays with energies above
$100\,\rm MeV$. Similar estimates have been given by Dorfi (1991a,b).
For the same SNR parameters, and assuming a differential
energy spectrum inside the remnant
proportional to $E^{-2.1}$ or $\alpha=4.1$, the flux
in the high energy $\gamma$-ray region is
$$
\eqalign{
F(>E)\approx 9\times10^{-11} \theta  &
\left(E\over1\,\rm TeV\right)^{-1.1}
\left(E_{\rm SN}\over 10^{51}\rm \,erg\right) \cr
& \left(d\over 1\,\rm kpc\right)^{-2}
\left(n\over\rm1\,cm^{-3}\right)\rm\, cm^{-2}s^{-1} \cr}
\eqno\autnum
$$
The surface brightness is
highest approximately at the transition from the free-expansion phase
to the Sedov phase. For fixed ejecta mass the radius at which this
occurs scales as $n^{-1/3}$ so that, combining this with the earlier
estimate of the total luminosity, we deduce that the maximum surface
brightness should scale with the external density as $n^{5/3}$.

\begfigwid 18 cm
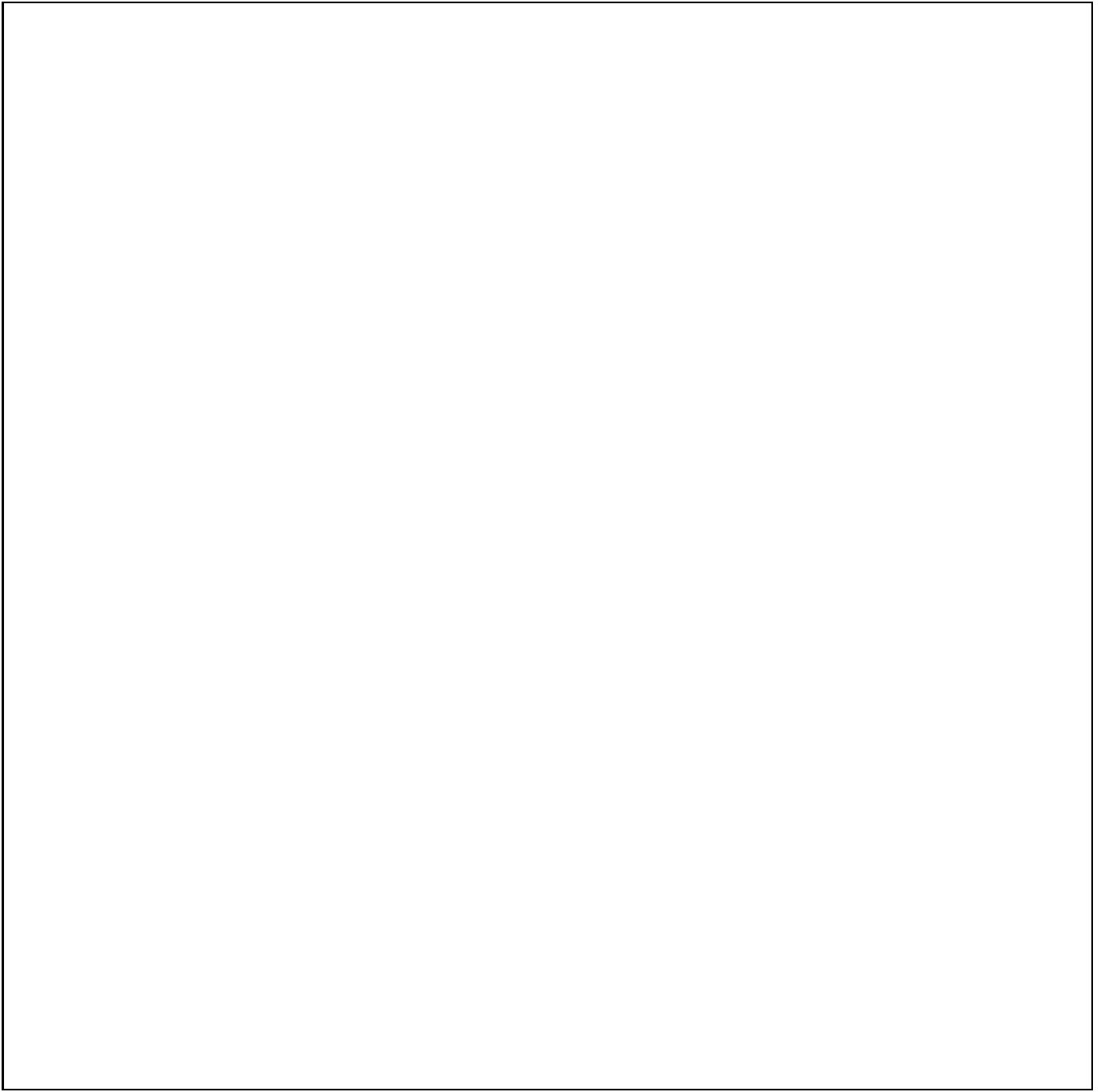
\figure{3}{As Fig.~1, but for an injection model which places most of the
injection and acceleration at late times (see DMV for details).}
\endfig

In Fig.~2 results are shown for an explosion in a higher density medium
with $n=5$ which confirm the above estimates (note that the neglect of
cooling is more serious in this case).
Although SNRs in denser regions of the ISM are stronger $\gamma$-ray
sources than those in low density regions, they are probably not as easy
to detect because of the problems associated with background subtraction and
source confusion. Indeed the COS-B measurements show that the diffuse
$E\ge100\,\rm MeV$ $\gamma$-ray emission at $\left|b\right|<5^\circ$ is
about $5\times10^{-4}$ and $1.5\times10^{-4}\,\rm ph\,cm^{-2}s^{-1}sr^{-1}$
from
the inner and outer parts of the galactic plane (Bloemen, 1987).
Thus the expected flux of
$\gamma$-rays within $1^{\circ}$ (the typical size of the near-by SNRs)
is about $(1.5 - 5)\times 10^{-7}$ making the detection of any source
in the galactic plane with an intensity less than
$10^{-7}\,\rm ph\,cm^{-2}s^{-1}$ hard even for the EGRET experiment
(see Fig.~4).

It is interesting to ask whether, by allowing the injection to be weaker
in strong shocks (as may be the case), it is possible to reduce the
surface brightness to the point where SNRs would be hard to detect.
In DMV a model of Mach number dependent injection was discussed which
has the property that most of the acceleration occurs towards the end of
the Sedov phase. In Fig.~3 the results for this model ($\alpha_I=1$ and
$\epsilon=0$) are presented for $n=0.1$. As expected the final total
luminosity (which is determined essentially by the explosion energy and
the ambient density if the acceleration is efficient) is not significantly
affected, but the surface luminosity is reduced because of the larger
remnant size.

Before turning to the observational implications of these models we
note that they do assume expansion of the SNR into a
uniform medium. While this is probably a good assumption for SNae of
type Ia, in other cases we would expect the ambient medium to be
perturbed by the stellar wind and photon flux of the SN precursor.
In detail this has not been explored, but qualitatively we expect that
there will be rather little CR production at early times as the young
SNR expands into a low-density interstellar bubble of radius $\le 10
\,\rm pc$ (although what acceleration there is may be to energies above
$10^{14}\,\rm eV$; {\it cf} V\"olk and Biermann, 1988).
Roughly two cases may have to be distinguished depending on the precursor
star. If the SNR has swept up a wind mass which is larger or of the same
amount
as the SN ejecta before reaching the edge of the bubble, then the internal
energy density and the pressure are roughly uniform over the SNR and
therefore `diluted' in the large bubble volume. With this postshock pressure
the outer shock will now shock heat the ISM matter
swept up and compressed by the
wind (and expanding HII-region), increasing the CR production and
$L_{\gamma}$. However $\theta$ may not reach the value which would be
possible if the ambient ISM had not been perturbed. In the second case, for
smaller mass of the precursor star, the wind mass is considerable smaller
than the ejecta mass. When the latter masses reach the edge of the bubble
and shock heat the swept-up wind shell,
the subsequent evolution should be very
similar to that of a SNR expanding into a uniform medium. The main effect
is thus probably to delay any significant CR production (and associated
$\gamma$-ray production) until the SNR has reached the edge of the
wind cavity.

\titlea{Discussion}

These results suggest that it should be possible to detect
$E_\gamma > 100\,\rm MeV$ $\gamma$-rays from at least some of
the relatively near-by ($d \leq 1 \rm \,
kpc$) SNRs, especially at high galactic latitudes where the background
confusion is less. It is interesting to note that claims
have been made for a detection of the north polar spur
(thought to be an old near-by remnant) in the COS-B and
\hbox{SAS-II} data at flux levels similar to those discussed here,
but the significance level is very low and the results depend
crucially on the background subtraction model used (Lebrun and Paul, 1985;
Bhat et al, 1985).

The other, perhaps more promising, possibility to test the SNR origin of
cosmic rays, is to look for very high energy $\gamma$ rays from these objects.
If, as the models suggest, the accelerated particle
spectra in SNRs are  hard  ($\alpha \sim 4$) and extend to at least
$10^{14} \,\rm eV$, then the fluxes of TeV $\gamma$-rays expected
are well above the sensitivities of modern Cherenkov telescopes,
even for SNR located in low density regions of the ISM.
Equation (9) shows that,
for $\theta E_{\rm SN}=10^{50}\,\rm erg$,
the flux of TeV $\gamma$-rays expected from a near-by SNR  (at $d \leq
1 \rm\, kpc$) exceeds  $10^{-12}\rm \, ph\,cm^{-2}s^{-1}$ at
$n \geq 0.1\rm \, cm^{-3}$.

Because the diffuse CR in the ISM have a softer spectrum than that
expected for SNR source spectra the diffuse $\pi^0$-decay background
should be much less of a problem at TeV energies and beyond than at
$100\,\rm MeV$. There is also a diffuse Galactic $\gamma$-ray background
resulting from IC scattering of the 2.7~K cosmic microwave background
by the CR electrons.  But at TeV energies this also
appears to be relatively less important than at $E\approx 100\,\rm MeV$
(see Appendix B).

In fact for ground-based detectors, which are the
only ones available at energies $E\ge 0.1\,\rm TeV$, the principal
background flux is that of the cosmic ray electrons themselves
because they produce atmospheric showers which are indistinguishable
from $\gamma$-ray induced showers of the same primary energy
\fonote{Of course this is not a problem  for
spacecraft instruments which can reject charged
particle events by using an anti-coincidence shield.}.
Although the background showers induced by
the nucleonic component of the CR are much more frequent, it is
possible to discriminate, at least partially,
between hadronic and electromagnetic
showers using various characteristics of the shower secondaries.
At TeV energies morphological analysis of the Cherenkov light
image has been shown to be a powerful technique for rejecting
hadronic showers, and at $E\ge 100\,\rm TeV$ the muon content
can be used. The sensitivities of modern ground-based detector
systems to cosmic $\gamma$-ray fluxes are mainly determined by the
efficiency of hadron rejection, which has to be estimated by
detailed Monte Carlo simulations.

The observational possibilities for detecting $\gamma$-rays
from SNRs are summarized in Fig.~4. At low $\gamma$-ray energies,
$E\ge 100\,\rm MeV$,  the typical range of the diffuse COS-B fluxes
from the galactic disc ($\left|b\right|\le 5\degr$) are shown,
bracketed from above by the average emission from the inner Galaxy
and from below by the slightly harder average emission from the
outer Galaxy (Bloemen, 1987). We plot the flux that would be observed
by an instrument with an acceptance of $1\degr$, roughly
corresponding to the expected angular size of SNRs. The three upper
curves are calculated spectra from SNR sources using our acceleration
models for three source spectra, $f(p)\propto p^{-\alpha}$ with
$\alpha = 4.1$, 4.2 and 4.3, and using the emissivities from
Appendix A. The spectra are normalized such that
$$
A = \theta \left(E_{\rm SN}\over 10^{51}\,\rm erg\right)
\left(n\over 1 \,\rm cm^{-3}\right)
\left(d\over 1\,\rm kpc\right)^{-2} = 1
\eqno\autnum
$$
where the quantities are as defined in section 2. For
$A\not=1$ the plotted source spectra should be scaled by a factor
$A$.  Remarkably these curves roughly coincide at $\gamma$-ray energies
$\le 1\,\rm GeV$, ie in the EGRET range.

\begfigwid 22 cm
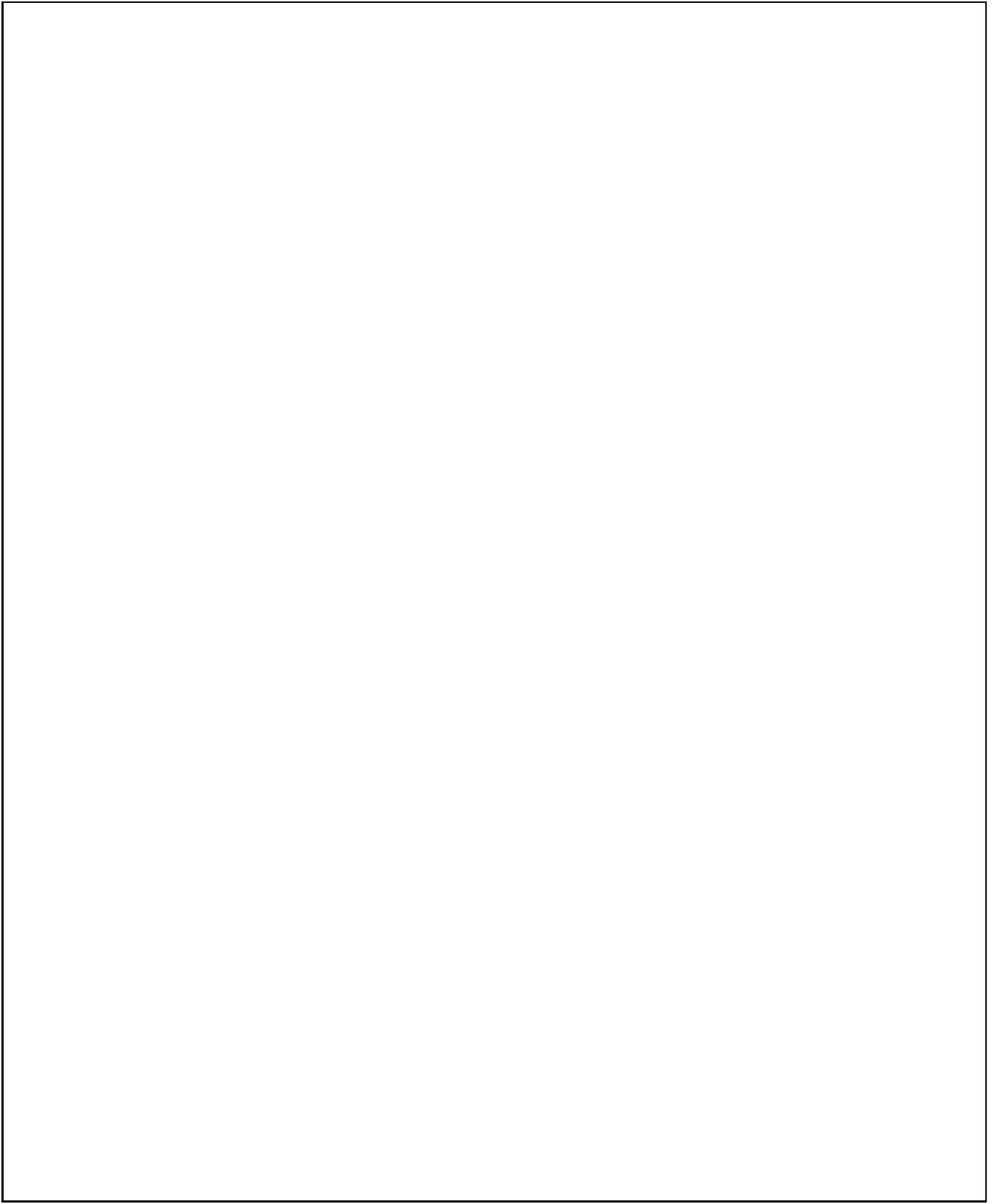
\figure{4}{The predicted $\gamma$-ray fluxes from a SNR together with
the various Galactic backgrounds (for a $1\degr$ source) and current
instrumental sensitivities; see text for details.}
\endfig

The diffuse $\gamma$-ray
emission, again within $1\degr$, for a total hydrogen column
density (both atomic and molecular)
$\left<N_{\rm H}\right>=\left<N_{\rm H_2}\right>+\left<N_{\rm HI}\right>
= 10^{22}\,\rm cm^{-2}$, a value typical of the outer Galaxy,
and a cosmic ray confinement volume of
size $L=10\,\rm kpc$, are given by the two lower curves.
The curve labeled $\pi^0$ corresponds to a diffuse Galactic cosmic
ray spectrum of the form
$J_p(E)\propto \left(E+m_pc^2\right)^{-2.75}$ (see section 2)
\fonote{While the spectrum above $10\,\rm GeV$ is well determined,
at lower energies, where solar modulation is dominant, we do not
really know the form of the demodulated ISM spectrum. This
particular form is compatible with the observations, and has
been extensively used by many authors in the past (eg
Stecker, 1971), but otherwise has no special significance.}.

The curve marked IC corresponds to the sum of the diffuse IC gamma
ray emissions as discussed in Appendix B and above. Finally the
background equivalence of the high energy diffuse interstellar
electrons is given by the dash-dotted line, based on the observations
of Nishimura et al (1980, 1990; see Appendix B).

The four hatched curve segments above $0.1\,\rm TeV$ correspond
to the $5\sigma$ sensitivities of different ground-based detection
systems for one year of operation ($\approx 100\,\rm h$ and
$\approx2000\,\rm h$ source exposure for a single imaging telescope and
charged particle detector arrays respectively). The sensitivity
(defined as the minimum flux of a $\gamma$-ray source detectable at
a level of confidence of $m$ standard deviations) is
$$
F_\gamma^{\rm (min)} (\ge E)\approx
m {P^{1/2}_{{\rm CR}\to\gamma} S^{1/2}_{\rm CR}\over
P_{\gamma \to \gamma} S_\gamma}
\left(J_{\rm CR}(\ge E)\Delta\Omega\over t\right)^{1/2}
\eqno\autnum
$$
where $J_{\rm CR}(\ge E)$ is the integral cosmic ray flux,
$\Delta\Omega\approx\pi\Delta\phi^2$ where $\Delta\phi$ is the
angular resolution of the detector, $S_\gamma$ and $S_{\rm CR}$
are the effective detection areas for photons and cosmic rays
respectively (in general the two are not equal),
$P_{{\rm CR}\to \gamma}$ is the rejection factor for background
events (the probability that a CR event will be misclassified as
a $\gamma$-ray event) and $P_{\gamma\to\gamma}$ is the probability
of correct classification of $\gamma$-ray induced showers.
The angular and lateral distribution of the Cherenkov radiation
as well as the muon content are characteristics which allow a
rather efficient separation of hadronic and electromagnetic
showers, $P_{{\rm CR}\to\gamma}\ll 1$ and $P_{\gamma\to\gamma}
\approx 1$ ({\it eg} Weekes, 1992). The high efficiency of the
Cherenkov imaging technique has been demonstrated recently by the
Whipple collaboration (Vacanti et al, 1991).

It should be noted that the detector sensitivities shown in Fig.4
correspond to point sources, namely to sources with angular size
smaller than $\Delta\phi$, where $\Delta\phi \simeq 0.1\degr$
corresponds to the resolution of modern imaging Cherenkov telescopes. For
extended sources the high resolution may not be of great help as far as
nucleon rejection is concerned. In particular, in case of the imaging
Cherenkov technique the `orientational' parameters of Cherenkov
images become rather ineffective, and the separation of proton- and
$\gamma$-induced showers should be done mainly through exploiting
the `shape' parameters . For a single telescope, the `shape'
parameters alone will provide
a sensitivity which is  a factor $2$ or $3$ lower than that shown in Fig.4
(e.g. Reynolds et al, 1993).
In fact, a system of imaging Cherenkov telescopes
will allow a much more effective
rejection of the CR background than a single telescope. The probability of
detection of the $\gamma$-ray induced shower by a single telescope
up to distances
$R \simeq 100 \, \rm m$ is close to 1, whereas the detection probability
of CR protons near the effective energy threshold of $\gamma$-rays is rather
low,  namely $\leq 0.1$. Therefore the requirement of detection of a shower
by a system of several ($\geq 3$) telescopes,separated by
$50-100\, \rm m$ from each other,practically  does not reduce  the number of
detected $\gamma$-rays, but it "suppresses" the CR background by factor of at
least $10^{-2}$, i.e. at the effective energy threshold of $\gamma$-rays
$S_{CR} \leq 10^{-2} S_{\gamma}$. Moreover, the software analysis of
the shape  of the Cherenkov images in different telescopes
(different projections) will be of additional help for the improvement
of the efficiency of $\gamma$/$p$ separation (Aharonian et al.,1993).
These two factors together allow us to expect that
future systems of imaging Cherenkov telescopes will approach an almost
`background-free' detection (i.e. the number of detected CR is smaller than
the number of detected $\gamma$-rays) down to $\gamma$-ray fluxes
$F_{\gamma} \sim 10^{-12} ph/cm^{2} s$, before the statistics of $\gamma$-rays
becomes the limiting factor, even for extended sources. Nevertheless, the
angular size of the $\gamma$-ray sources for a reasonable value of the
aperture of the imaging camera ($\leq 5\degr$) should not exceed
$\sim 2\degr$, since beyond $2\degr$ the efficiency of detection of
$\gamma$-rays drops sharply. These results will be published elsewhere.

\par
 As we will show below the radius of
a SNR only exceeds $10\,\rm pc$ after a minimum age of a few
thousand years, provided we limit ourselves to ambient gas densities
$n\ge0.1\,\rm cm^{-3}$. Obviously a SNR in a lower density environment
would be very difficult to detect. Moreover for more favourable
parameters, eg those characteristic of Tycho's SNR, the
time when the source size reaches $10\,\rm pc$ is approximately
$2\times10^4\,\rm y$. Note that a radius of $10\,\rm pc$ implies
an angular size $\le 1\degr$ if the source is located at more
than $1\,\rm kpc$. On the other hand the $\gamma$-ray luminosity
reaches a plateau level at $t_1\le 10^4\,\rm y$, and continues
to be at this level during several $t_1$, depending on the
parameters (see Fig.~1). Since $F_\gamma\propto L_\gamma/d^2$
the best candidates for detection are SNRs with age of order
$10^4$ years at a distance $d\approx 1\,\rm kpc$. Of course
for SNRs in dense environments the $\gamma$-ray fluxes may be considerable
even for $d> 1\,\rm kpc$.
In galaxies similar to ours the core collapse SNe form massive stars
outnumber SN type Ia by almost an order of magnitude
(Evans, van den Bergh, and McClure, 1989;
Tammann, 1992), and such stars form typically as clusters in dense
clouds. Given the influence of these stars on the ambient medium, it is
nevertheless not clear how frequent SNRs in a $n \geq 10 \, \rm cm^{-3}$
medium are. Still, the enhanced SNR rate in the molecular ring of the Galaxy
at $d \simeq 4 \, \rm kpc$ (eg Bloemen et al., 1993)
make this region in principle very attractive
for the detection of distant SNRs. In the 100 MeV region such detection
appears practically quite difficult
as becomes clear from the fact that the COS-B diffuse
galactic fluxes within $1\degr$ (a few $10^{-7}\,\rm
cm^{-2}s^{-1}$, see Fig.~4) approximately correspond to
the sensitivity of the EGRET detector for discrete $\gamma$-ray
sources in the galactic disc.

As can be seen from Fig.~4, the situation is better at TeV energies,
where $\gamma$-rays from SNRs may be detected up to $d\approx 10\,\rm kpc$
provided $n\ge 1\,\rm cm^{-3}$. (It is interesting to note that for
SNRs with marginally detectable fluxes and angular sizes of
order $1\degr$ a lowering of the energy threshold below $1\,\rm TeV$
is not necessarily the most efficient detection strategy because of the
rising electron background.) Therefore future imaging Cherenkov
telescopes (ICT) will be very useful in searching for high energy
$\gamma$-rays from SNRs within wide ranges of ages and distances.
The detection of near-by ($d<1\,\rm kpc$)
SNRs is complicated by the large angular size of the
sources, but, as was mentioned above,
possible with arrays of ICTs.
The Whipple GRANITE detector (Akerloff et al, 1990)
consisting of two $\approx 10\,\rm m$ imaging telescopes has been
built recently, and the HEGRA imaging telescope array (Aharonian
et al, 1991) consisting of five ICTs is now under construction
at La Palma. The high angular resolution of ICT arrays will allow
the $\gamma$-ray spectrum to be measured with rather high accuracy
($\Delta E/E\approx 20\%$) and thus give direct information
about the spectrum of accelerated particles in the SNR. The
weak dependence of the fluxes at 0.1 to $1\,\rm GeV$ on the shape of
spectrum of accelerated particles (assuming a power-law form;
see Fig.~4) means that measurements in this energy range are
in some ways less informative than those at TeV energies.
However, for the same reason, EGRET measurements can provide
a rather model-independent estimate of the normalization
parameter $A$ [Eq.~(9)] and thus the total nonthermal power of
SNRs.

The important practical question of identifying candidate SNRs
for observation can best be illustrated by an example. Let us
consider Tycho's SN of 1572 which is at an estimated distance of
$d=2.25\pm0.25\,\rm kpc$ (e.g.~Heavens, 1984). From the historical
light curve the SN is commonly assumed to have been of type Ia,
an accreting white dwarf system, for which our model assumptions of
a uniform ambient medium are probably well justified ({\it cf} the discussion
at the end of section 2). According to the most recent radio observations
(Tan and Gull, 1985) the expansion parameter $t\dot R/R = 0.462\pm0.024$
where $R(t)$ is the outer SNR radius and $\dot R(t)$ its time derivative
at time $t$ after the initial explosion. Thus the expansion parameter has
decreased quite far towards its asymptotic value of 0.4 in the Sedov phase.
This dynamical argument is probably more reliable than present X-ray
models which need to take account of non-equilibrium electron temperatures
and non-equilibrium ionisation states behind the outer shock as well
as nonsolar chemical composition in the ejecta which are part heated by
the reverse shock.  Thus Tycho's SNR is probably quite close to the
Sedov phase. Different X-ray models give results ranging from
$n=0.28\,\rm cm^{-3}$ and $E_{\rm SN}=7\times 10^{50}\,\rm erg$
(Hamilton et al, 1986) to
$n=1.13\,\rm cm^{-3}$ and $E_{\rm SN}=1.9\times 10^{50}\,\rm erg$
(Smith et al, 1988).
A value of $E_{\rm SN}/n = 0.17\times 10^{51}\,\rm erg\,cm^3$ with
a quoted error of about 20\% from the latter reference agrees reasonably
well with the dynamical value of
$E_{\rm SN}/n = 0.2\times 10^{51}\,\rm erg\,cm^3$
inferred by Heavens (1984) from radio and X-ray data
and gives the value $n E_{\rm SN} = 0.2\times 10^{51}\,\rm erg\,cm^{-3}$.
Thus assuming $\theta\approx 0.2$ we have $A\approx 0.01$. If the Sedov
phase is far from being reached, then we may have $\theta\ll 0.2$ and
$A\ll 10^{-2}$; however it would be hard to reconcile this with the
amounts of shock heated gas seen in the X-ray band as well as the kinematical
evidence from the expansion parameter.  Alternatively, it may be that
in this particular remnant particle acceleration is inefficient and
$\theta \ll 0.2$; but then, if the Galactic SNRs are the sources of the
cosmic rays, the remaining SNRs must be even stronger accelerators
and we soon run into energy problems. Thus, on balance, the prediction of
$A\approx 0.01$ for Tycho's SNR is a hypothesis worth testing.

The spectrum inside such a young SNR is presumably very hard with
$\alpha\le 4.1$. This is due to the fact that the shock is still very
strong and wave heating is inefficient (DMV) so that even in a simple
test particle approximation the spectrum is very hard (V\"olk, Zank and
Zank, 1988). Nonlinear effects tend to flatten the high-energy end of the
spectrum so that assuming a spectral slope of $4.1$ for Tycho is not
unreasonable. On looking at Fig.~4 we see that an EGRET detection of
Tycho appears impossible, but that it should be visible at $1\,\rm TeV$
for imaging Cherenkov arrays and at $20\,\rm TeV$ for a $1\,\rm km^2$
AEROBICC array. The present radius of Tycho's SNR corresponds to
about 4 arc minutes.

This situation should persist for $10^4$ years at which epoch Tycho's
SNR will have an angular radius of about $0.25^\circ$.  The fraction
of the explosion energy converted to cosmic ray energy should remain
at $\theta\approx 0.2$ or higher while the spectral slope will begin
to increase from $\alpha\approx 4.1$ to $\alpha\approx 4.3$ only towards
the end of the evolution. Therefore a remnant of this type may be
better observed at somewhat later times.  It is worth noting that if
Tycho were located at $d=1\,\rm kpc$ we would have $A\approx 0.1$ and
detection would be much simpler. Obviously SNRs in denser environments
are better candidates at the same distance and comparable evolutionary
epochs. But we are dealing here with the statistics of a small number
of nearby sources and each candidate requires an individual evaluation
before sensitive survey instruments become available.

Most models of CR acceleration in SNRs predict upper cutoffs
in the spectrum between $10^2$ and $10^3\,\rm TeV$. There will however
be some SNRs whose precursor stars were massive OB stars or Wolf-Rayet stars.
These SNae explode into a medium
where the magnetic field is that in the wind of the
progenitor star and under these circumstances energies as high
as $10^3$ or $10^4\,\rm TeV$ may be possible (V\"olk and Biermann, 1991).
If such special sources have power-law spectra a clear observational
test of their existence is possible with the existing UMC (Utah, Michigan,
Chicago) detector (Fick et al, 1991). The $5\sigma$ sensitivity of this
system above $100\,\rm TeV$ at present reaches about
$3\times10^{-14}\,\rm ph\,cm^{-2}s^{-1}$ in one year of operations
(see Fig.~4). Therefore the efficiency of CR acceleration in SNR
to energies significantly above $10^3\,\rm TeV$ may soon be
tested for all near-by SNRs using the all sky survey of the UMC detector.

The new generation air shower detectors like CRT (the Cosmic Ray
Tracking experiment; Heintze et al, 1989)
MILAGRO (Multiple Image Los Alamos Gamma Ray Observatory;
Williams et al 1991) and the $1\,\rm km^2$ AIROBICC
(AIR shower Observation By angle Integrating Cherenkov Counters;
Lorenz, 1992) could observe SNR sources above $10\,\rm TeV$ provided the
source CR spectra extend to energies $\ge 10^2\,\rm TeV$
(note that the spectra in Fig.~4 have been drawn with no high energy
cutoff). Thus even negative results, in the form of firm upper limits
in the $E\ge 10\,\rm TeV$ range, could be of great importance in
determining the upper cutoff of the acceleration process in SNRs.

Although this paper concentrates, for obvious reasons, on the $\gamma$-ray
visibility, there is another possibility to detect high energy
stable neutral particles from SNRs. From Table A1 we expect
(for hard source spectra, $\alpha\approx2$) a neutrino
flux above $1\,\rm TeV$ of
$$
F_{\nu_\mu}(\ge 1{\,\rm TeV}) \approx
10^{-10} A \,\rm neutrinos\,cm^{-2}s^{-1}
\eqno\autnum
$$
Such fluxes may be marginally detectable for DUMAND (Deep Underwater
Muon And Neutrino Detector; eg Okada, 1991) in
the case of near-by SNRs. This is especially interesting because of
the all sky survey capability of DUMAND.

Besides SNRs in our own Galaxy, it may be
possible to use the same combination of techniques to detect
the integrated $\gamma$-ray flux of some favourable near-by
normal galaxies. Such galaxies do not have an active
nucleus, but presumably contain the same types of cosmic ray
sources as our Galaxy. Examples might be the Magellanic Clouds
and the near-by starburst galaxy M82. This will be discussed in
a separate paper.

\titlea{Summary}

Recent nonlinear particle acceleration models for SNRs have been
used to predict their $\gamma$-ray luminosities. A detailed analysis
of the various backgrounds shows that the most favourable energy
range for detecting $\gamma$-rays from SNRs is probably from 1 to
$10\,\rm TeV$.  In addition, to characterise the spectra, detections
in the GeV region and detections or upper bounds in the PeV region
would be very useful and appear possible with current detectors.

Galactic SNRs should be detectable $\gamma$-ray sources at distances of
up to a few kpc in regions of the ISM where the mean density is of
order $1\,\rm cm^{-3}$ or more, if, as is generally supposed, they are
the main sources of the cosmic rays below $10^{14}$ to $10^{15}\,\rm eV$.
Even the detection of one or two SNR which are not dominated by a
pulsar, at the level suggested in this paper would be a convincing
demonstration of the SNR origin of cosmic rays.

\acknow{Various aspects of this work were discussed at meetings in Vulcano
and Ringberg. The authors are especially indebted to E.Dorfi,
A.Heavens,R.Tuffs, and  R.Wielebinski
for helpful remarks, and to P.O.Lagage for critical
comments on manuscript.}

\begref{References}
\ref
Aharonian F A,
Akhperjanian A G,
Kankanian R S,
Mirzoyan R G,
Samorski M,
Stamm W,
Bott-Bodenhausen M,
Lorenz E,
Sawallisch P,
1991, Proposal for Imaging Air Cherenkov
Telescopes in the HEGRA Particle Array, Kiel
\ref
Aharonian F A,
Chilingarian A A,
Mirzoyan R G,
Konopelko A K,
Plyaskeshmikov A V,
1993,
Experimental Astronomy (in press)
\ref
Aharonian F A, 1991, Ap\&SS 180, 305
\ref
Akerloff CW, Cawley M F, Feegan D J, Hillas A M, Lamb R C,
Lewis D A, Meyer D I, Weekes T C, 1990, Nucl. Phys. B
(Proc. Suppl.) 14A, 237
\ref
Berezhko E G,  Krymsky G F, 1988, Usp. Fiz. Nauk. 154, 49
(English translation, Sov. Phys. Usp. 31, 27)
\ref
Bhat et al, 1985, Nature 314, 515.
\ref
Blandford R D,  Eichler D, 1987, Phys. Rep. 154, 1
\ref
Blandford R D, Ostriker J P, 1980, ApJ 237, 793
\ref
Bloemen J B G M, 1987, ApJ, 317, L15
\ref
Bloemen, J B G M, Dogiel, V A, Dorman V L, Ptuskin V S, 1993,A\&A 267, 372
\ref
Blumenthal G R, Gould R J, 1970, Rev. Mod. Phys. 72, 237
\ref
Bogdan T,  V\"olk H J, 1983, A\&A 122, 129
\ref
Cronin J W, 1992, Nucl. Phys. B (Proc. Suppl.) 25A, 137
\ref
Dermer C D, 1986, A\&A 157, 223
\ref
Dorfi E A, 1990, A\&A 234, 419
\ref
Dorfi E A, 1991a, Proc 22nd ICRC (Dublin) 1, 109
\ref
Dorfi E A, 1991b, A\&A 251, 597
\ref
Drury L O'C, Markiewicz W J,  V\"olk H J [DMV], 1989,
A\&A 225, 179
\ref
Evans R S, van den Bergh S, McClure R D, 1989. ApJ 345, 752
\ref
Fick B E, Borione A, Covault C E, Cronin J W, Gibbs H A, Krimm H A,
Mascarenhas N C, McKay T A, Muller D, Newport B J, Ong R A, Rosenberg B J,
1991, Proc 22 ICRC (Dublin) 2, 728
\ref
Gaisser T K, 1990, Cosmic Rays and Particle Physics, Cambridge University
Press, Cambridge.
\ref
Hamilton A J S, Sarazin C L, Symkowiak A E, 1986,
ApJ 300, 713
\ref
Heavens A F, 1984, MNRAS 211, 195
\ref
Heintze J et al, 1989, Nuc. Instr. and Methods A227, 29
\ref
Higdon J C,  Lingenfelter R E, 1975, Astrophys. J. 198, L17
\ref
Jokipii J R, Ko M, 1987, Proc 20 ICRC (Moscow) 2, 179
\ref
Jones L W, 1990, Proc 21st ICRC (Adelaide) 2, 75
\ref
Jones T W, 1993, ApJ (submitted)
\ref
Juliusson E, Meyer P, M\"uller D, 1972, Phys. Rev. Lett. 21, 445
\ref
Kang H, Drury L O'C, 1992, ApJ 399, 182
\ref
Kang H, Jones T W, 1991, MNRAS 249, 439
\ref
Krymsky G F, Petukhov S I, 1980, Pis'ma Astron. Zh. 6, 227
(English translation, Sov. Astron. Lett. 6, 124)
\ref
Lagage P O, Cesarsky C J, 1983, A\&A 125, 249
\ref
Lebrun F,  Paul J, 1985, Proc 19 ICRC (la Jolla) 1, 309.
\ref
Lorenz E, 1992, Presentation at the Paris workshop on VHE
$\gamma$-ray detectors
\ref
Markiewicz W J, Drury L. O'C., V\"olk H J, 1990, A\&A 236, 487
\ref
Moraal H,  Axford W I, 1983, A\&A 125, 204
\ref
Nishimura J, Fujii M, Taira T, Aizu E, Hiraiwa H, Kobayashi T,
Niu K, Ohta I, Golden R L, Koss T A, Lord J J,  Wilkes R J,
1980, ApJ 238, 394
\ref
Nishimura J, Fujii M, Kobayashi T, Aizu H, Komori Y,
Kazuno M, Taira T, Koss T A, Lord J J, Wilkes R J,  Woosley J,
1990, Proc 21st ICRC (Adelaide) 3, 213.
\ref
Okada A, 1991, in: Astrophysical Aspects of
the Most Energetic Cosmic Rays, eds. Nagano M and Takahara F,
World Scientific, Singapore, p.483
\ref
Prischep V L, Ptuskin V S, 1981, Astron. Zh. 58, 779
(English translation, Sov. Astron. 25, 446)
\ref
Reynolds P T, Akerlof C.W., Cawley M F, Chantel M, Fegan D J, Hillas A M,
Lamb R C, Lang M J, Lawrence M A, Lewis D A, Macomb D, Meyer A I,
Mohanty G, O'Flaherty K S, Punch M, Schubnell M S, Vacanti G,
Weekes T C, Whitaker T, 1993,ApJ 404, 206
\ref
Smith A, Davelaar J, Peacock A, Taylor B G, Morini N, Robba N R, 1988
ApJ 325, 288
\ref
Stecker F W, 1971, Cosmic Gamma Rays (NASA Scientific and Technical
Information Office), NASA SP-249.
\ref
Stecker F W, 1979, ApJ 228, 919
\ref
Swordy S P, M\"uller D, Meyer P, L'Heureux J, Grunsfeld J M, 1990,
ApJ 349, 625
\ref
Tammann G A, 1992, in: Supernovae, Bludman S, Mochkovitch R and
Zinn-Justin (eds.), Les Houches, Elsevier Science Publishers B.V., p.1
\ref
Tan S M, Gull S F, 1985, MNRAS 216, 949
\ref
Vacanti G, Cawley M F, Colombo E, Fegan D J, Hillas A M, Kwok P W,
Lang M J, Lamb R C, Lewis D A, Macomb D J, O'Flaherty K S, Reynolds P T,
Weekes T C, 1991, ApJ 377, 467
\ref
V\"olk H J, Biermann P L, 1988, ApJ 333, L65
\ref
V\"olk H J, Drury L O'C, Dorfi E A, 1985, Proc 19th ICRC (La Jolla)
3, 148
\ref
V\"olk H J, Zank L, Zank G, 1988, A\&A 188, 274
\ref
Weekes T, 1992, Space Sci. Rev. 59, 315
\ref
Williams D A, et al,
1991,
Proc 22nd ICRC (Dublin) 2, 684
\endref

\appendix{A: The emissivity of very high energy neutral particles in
p-p interactions}

On the assumption that the primary protons have a power-law
differential energy spectrum,
$$
N_{\rm p}(E) = N_0 E^{-\alpha + 1},\eqno{\rm A1}
$$
the emissivity for the secondary particles produced in p-p interactions
may be written in the form
$$
{\cal E}_S(E) = \left<mx\right>_S^\alpha\sigma_{pp}c N_{\rm p}(E)
\eqno{\rm A2}
$$
where $\sigma_{pp}$ is the inelastic p-p cross-section,
$c$ is the velocity of light and
$$
\left<mx\right>_S^\alpha = \int_0^1 x^\alpha g(x)\,dx
\eqno{\rm A3}
$$
is the so-called spectrum-weighted moment of the inclusive cross-section
(see, e.g.~Gaisser 1990; $x$ is the ratio of the secondary particle energy
to that of the primary proton and $g(x)$ is the dimensionless inclusive
cross-section).

A comprehensive discussion of the spectrum-weighted moments for secondary
hadrons based on the interaction of accelerator beams with fixed targets
at beam energies $\le 1 \,\rm TeV$ has been presented by Gaisser (1990).
In table A1 we present the spectrum-weighted moments for $\pi^0$-mesons from
Gaisser's book.

As the production and decay of secondary $\pi^0$ mesons is the main source of
$\gamma$-rays (the contribution from the next most important channel via
$\eta$ mesons is only about $\left<mx\right>_\eta/\left<mx\right>_{\pi^0}
\approx \left(m_{\pi^0}/m_\eta\right)^2\approx 0.06$)
$$
{\cal E}_\gamma(E)=
2\int_E^\infty{{\cal E}_{\pi^0}(E')\over E'}dE'
= {2\over\alpha+1}{\cal E}_{\pi^0}(E)
\eqno{\rm A4}
$$
and thus
$$
\left<mx\right>^\alpha_\gamma
\approx \left<mx\right>^\alpha_{\pi^0}{2\over\alpha+1}
\eqno{\rm A5}
$$

Though the values of $\left<mx\right>_{\pi^0}$ given in Table A1
are calculated for fixed target accelerator data with $E\le 1\,\rm TeV$
it is expected that they correctly characterise also the energy region beyond
$1\,\rm TeV$ (Gaisser, 1990). Indeed, the spectrum-weighted moments for gamma
rays calculated on the basis of data from p-\=p colliders at
$\sqrt s = 630\,\rm GeV$ (the CERN UA7 experiment) can be approximated in
simple form (A. Erlykin, personal communication) by
$$
\log\left<mx\right>^\alpha_\gamma
= 1.49 -2.73\alpha + 0.53 \alpha^2
\eqno{\rm A6}
$$
Comparison of Eq.~A6 with $\left<mx\right>^\alpha_\gamma
\approx \left<mx\right>^\alpha_{\pi^0}{2\over\alpha+1}$
and the values of $\left<mx\right>_\alpha^{\pi^0}$ in Table A1
shows agreement within 20\%, which seems very reasonable if account is
taken of the uncertainty associated with the contribution through
$\eta$-mesons, especially at high energies.

\begtabfull
\tabcap{A1}{Spectrum weighted moments for the production of secondaries in
p-p collisions.}
\halign{#\hfil&&\quad#\hfil\cr
\noalign{\hrule\medskip}
$\alpha$ & 1.0 & 1.2 & 1.4 & 1.6 & 1.8\cr
\noalign{\medskip\hrule\medskip}
$\left<mx\right>^\alpha_\gamma$ Eq.~(6) &
0.19 & 0.094 & 0.051 & 0.030 & 0.019 \cr
\noalign{\medskip\hrule\medskip}
$\left<mx\right>^\alpha_{\pi^0}$&\cr
Gaisser (1990) & 0.17 & 0.092 & 0.066 & 0.048 & 0.036 \cr
\noalign{\medskip\hrule\medskip}
$\left<mx\right>^\alpha_{\rm n}$ Eq.~(11) &
0.062 & 0.050 & 0.041 & 0.034 & 0.029 \cr
Gaisser (1990) &
0.05 & 0.05 & 0.04 & 0.03 & 0.03 \cr
\noalign{\medskip\hrule\medskip}
$\gamma\over\nu_\mu + \nu_{\bar\mu}$ Eq.~(9, 10) &
0.95 & 0.80 & 0.67 & 0.56 & 0.46 \cr
Gaisser (1990) & 0.98 & 0.86 & 0.77 & 0.66 & 0.58 \cr
\noalign{\medskip\hrule}}
\endtab

At p-p interactions, as a result of the decay of charged pions,
neutrinos will also be produced. From the point of view of detectability
the most interesting are the muonic neutrinos, $\nu_\mu$ and $\nu_{\bar\mu}$,
produced in the decays (1) $\pi^\pm\to\mu^\pm+\nu_\mu$ and (2)
$\mu^\pm\to{\rm e}^\pm+\nu_\mu+\nu_{\rm e}$. Taking into account
that
$$
{\cal E}_{\pi^0}\approx {1\over 2}{\cal E}_{\pi^\pm}
\eqno{\rm A7}
$$
and the relation (eg Stecker, 1979)
$$
{\cal E}^{(1)}_\mu(E) =
{1\over 2\eta}\int_{E/2\eta}^\infty{{\cal E}_{\pi^\pm}(E)\over E} dE
= {(2\eta)^\alpha\over\alpha+1}{\cal E}_{\pi^\pm}(E)
\eqno{\rm A8}
$$
gives for the first decay
$$
{\cal E}^{(1)}_{\nu_\mu} / {\cal E}_\gamma = (2\eta)^\alpha
\eqno{\rm A9}
$$
where $\eta=1-m_\mu^2/m_\pi^2$.
An exact calculation of the $\nu_\mu$ spectrum from decay (2) is more
complicated, however it can be estimated as follows. In charged pion
decays on average a fraction $(1-\eta)$ of the pion energy is transferred
to the muons, while approximately the same ($\approx 1/3$) part of the
muon energy goes to the decay products, e$^\pm$, $\nu_\mu$ and $\nu_{\rm e}$.
Thus in this simple approximation
$$
{\cal E}_{\nu_\mu}^{(2)}/{\cal E}_\gamma \approx
(\alpha+1)\xi^\alpha;\qquad \xi={1-\eta\over 3}\approx 0.026
\eqno{\rm A10}
$$
Equations A10 and A9 imply that both channels (1) and (2) make comparable
contributions to the $\nu_\mu$ production rate and that for hard proton
spectra ($\alpha\approx 1$) the ratio of gamma to neutrino production is
close to unity.

Over distances of order $(E/10^{17}\,\rm eV)\, kpc$ neutrons can also be
considered as quasistable neutral particles of possible astrophysical interest.
The emissivity of neutrons can be calculated using the mean neutron
multiplicity $\left<m\right>\approx 1/4$ and the normalised $x$-distribution
of neutrons,
$g(x) = 3 (1-x)^2$, suggested by Jones (1990) with the result
$$
\left<mx\right>^\alpha_{\rm n} =
{3\over 4}{\Gamma(3)\Gamma(\alpha+1)\over\Gamma(\alpha+3)}
= {3\over 2} {1\over(\alpha+3)(\alpha+2)(\alpha+1)}
\eqno{\rm A11}
$$
The various spectrum-weighted moments for stable neutral particles are
presented in Table~A1.

\appendix{B: The diffuse Galactic $\gamma$-ray emission at very high energies}
The Galactic $\gamma$-ray background consists of two components, $\pi^0$ and
IC, resulting from interactions of cosmic ray nucleons and electrons
with the ISM. The flux of the first component can be presented in the
form (Aharonian, 1991)
$$
J^{\pi^0}_\gamma =
{1\over4\pi}{\cal E}(>E)\eta\left<N_{\rm H}\right>
=\left<mx\right>^\alpha_\gamma\sigma_{\rm pp}\eta
J_{\rm CR}(>E)\left<n_H\right>
\eqno{\rm B1}
$$
where $\sigma_{\rm pp}$ is the inelastic p-p cross-section,
$\left<mx\right>^\alpha_\gamma$ is the spectrum weighted moment
for $\gamma$-rays (see Appendix A), $J_{\rm CR}(>E)$ is the integral
energy spectrum of the Galactic cosmic rays, $\eta$ is the contribution
of nuclei, mostly $\rm ^4He$, in the cosmic rays and the ISM to $\pi^0$
production (for standard composition $\eta\approx 1.5$, see eg Dermer, 1986)
and $\left<N_{\rm H}\right>=\left<N_{\rm HI}\right>+\left<N_{\rm H_2}\right>$
is the total hydrogen (atomic and molecular) column density in the given
direction.

Radio measurements in the $21\,\rm cm$ and $2.6\,\rm mm$ lines
provide rather accurate estimates of $\left<N_H\right>$, so the main
uncertainty is associated with the cosmic ray flux above $1\,\rm TeV$.
Indeed, if the CR proton spectrum is taken to be
$$
J_{\rm CR}(>E)\approx 7\times10^{-6}
\left(E\over 1\,\rm TeV\right)^{-1.75}\,\rm cm^{-2}s^{-1}sr^{-1}
\eqno{\rm B2}
$$
then at $E\gg 100\,\rm MeV$ we have
$$
J_\gamma(>E) \approx
1.2\times10^{-10}
\left(E\over 1\,\rm TeV\right)^{-1.75}
\left<N_H\right>_{22}
\,\rm cm^{-2}s^{-1}sr^{-1}
\eqno{\rm B3}
$$
whereas for the Hillas (1981) representation of the spectrum,
$$
J_{\rm CR}(>E)\approx 10^{-5}
\left(E\over 1\,\rm TeV\right)^{-1.6}\,\rm cm^{-2}s^{-1}sr^{-1}
\eqno{\rm B4}
$$
the corresponding diffuse $\gamma$-ray spectrum is
$$
J_\gamma(>E) \approx
2.2\times10^{-10}
\left(E\over 1\,\rm TeV\right)^{-1.6}
\left<N_H\right>_{22}
\,\rm cm^{-2}s^{-1}sr^{-1}
\eqno{\rm B5}
$$
where $\left<N_H\right>_{22} = \left<N_H\right>/10^{22}\,\rm cm^{-2}$.
Thus there is about a factor two uncertainty in the $\gamma$-ray flux above
$1\,\rm TeV$ associated with uncertainty in the Galactic CR flux.

The flux related to the electron component of the CR is even less
certain. However the observations reported by Nishimura et al
(1980, 1990) do extend to energies slightly above $1\,\rm TeV$
and show that the flux of high energy electrons is reasonably
represented by a power-law,
$$
J_{\rm e}(E) = 7.6\times 10^{-9}
\left(E\over 1\,\rm TeV\right)^{-3.3}
\,\rm cm^{-2}s^{-1}sr^{-1}TeV^{-1}
\eqno{\rm B6}
$$
in the region $10\,{\rm GeV}<E<1\,\rm TeV$. Here we will assume that
this spectrum extends to well above $1\,\rm TeV$.

Although at $100\,\rm MeV$ the contributions from brems\-strahl\-ung
and IC processes on starlight and $2.7\,\rm K$ cosmic microwave background
(CMB) are comparable, at VHE ($E\ge 1\,\rm TeV$) only IC on the CMB
remains important because for $\gamma$-ray energies
$E\ll 4 m_{\rm e}^2c^4/3kT\approx 10^3\,\rm TeV$ it works in the
Thomson limit and produces a hard spectrum of $\gamma$-rays.
The emissivity of this process in the case of a narrow-spectrum
background photon field with an average photon energy $w_0$ and
number density $n_0$ is (see, eg, Blumenthal and Gould, 1970)
$$
\eqalign{
{\cal E}_\gamma^{\rm IC} =
\pi r_0^2 c n_0 N_0 & 2^{\beta+3}
{\beta^2+4\beta+11\over(\beta+3)(\beta+1)(\beta+5)} \cr
& \left(w_0\over m_{\rm e}c^2\right)^{\beta-1\over 2}
\left(E\over m_{\rm e}c^2\right)^{-{\beta+1\over 2}}\cr}
\eqno{\rm B7}
$$
where
$$
N_{\rm e}(E) = {4\pi\over c}J_{\rm e}(E)
= N_0 E^{-\beta}
\eqno{\rm B8}
$$
is the spectrum of the relativistic electrons.

Thus, for IC of the high energy cosmic ray electrons on the CMB
with $w_0\approx 3kT$ and $n_0\approx 400 \,\rm cm^{-3}$ we have
$$
\eqalign{
&J^{\rm IC}_{\rm CMB}(>E)
= {1\over 4\pi} {\cal E}^{\rm IC}_{\rm CMB}(>E) L_{\rm e}\cr
& \qquad\approx
4\times 10^{-11}
\left(E\over 1\,\rm TeV\right)^{-1.15}
\left(L_{\rm e}\over 10\,\rm kpc\right)
\,\rm cm^{-2}s^{-1}sr^{-1} \cr}
\eqno{\rm B9}
$$
where $L_{\rm e}$ is the characteristic size of the confinement
region of CR electrons in the Galaxy (in the given direction).

Equation B7 shows that, in the Thomson limit,
$$
{\cal E}^{\rm IC}_\gamma \propto
n_0 w_0^{\beta-1\over 2}\propto W_0 w_0^{\beta-3\over 2}
\eqno{\rm B10}
$$
where $W_0=n_0 w_0$ is the energy density of the photon field.
Let us consider in this limit the IC emissivities for scattering
the CMB, optical and far IR radiation fields. The optical radiation,
with a mean photon energy of $1.5\,\rm eV$ is estimated to have an energy
density of $0.5\,\rm eV\,cm^{-3}$ and the far IR with a mean photon
energy of $0.01\,\rm eV$ to have an energy density of $0.2\,\rm eV\,cm^{-3}$.
Then in the Thomson limit
$$
{\cal E}^{\rm IC}_{\rm CMB}
\approx 0.15
{\cal E}^{\rm IC}_{\rm O},
\qquad
{\cal E}^{\rm IC}_{\rm CMB}
\approx 0.5
{\cal E}^{\rm IC}_{\rm IR}
\eqno{\rm B11}
$$
so that the $\gamma$-ray background due to the IC scattering of optical photons
is roughly three times that due to the scattering of far IR photons which in
turn is about twice that due to scattering of the CMB photons.
However, the parameter $\delta= E_\gamma w_0/m_{\rm e}^2 c^4$ is
$\approx 6(E_\gamma/1{\,\rm TeV})$
for optical photons and
$\approx 0.04(E_\gamma/1{\,\rm TeV})$ for IR photons.
The Thomson cross-section, $\sigma_{\rm T}$, is only appropriate for
$\delta\ll 1$ and for $\delta\ge0.01$ the Klein-Nishina cross-section,
$\sigma_{\rm KN}$, must be used. Asymptotically
$$
{\sigma_{\rm KN}\over\sigma_{\rm T}} =
\cases{ 1 - 2\delta^{1/2} & if $\delta\ll 1$\cr
{3\over 8}\delta^{-1/2}\left[1+\ln\left(2\delta^{1/2}\right)\right] &
if $\delta\gg 1$\cr}
\eqno{\rm B12}
$$
so that the cross-section drops by a factor of two at $\delta\approx0.05$ and
by
more than a factor 10 at $\delta\approx 5$. Thus at $1\,\rm TeV$ and above
the dominant contribution to the IC background is in fact the component
produced by the CMB (the spectra also steepen in the Klein-Nishina region).

It is also easy to show that the electron bremsstrahlung contribution is
negligible when compared to the IC and $\pi^0$ backgrounds,
$$
J_\gamma(>E)\approx 3.8\times10^{-13}
\left(E\over 1\,\rm TeV\right)^{-2.3}
\left<N_{\rm H}\right>_{22}
\,\rm cm^{-2}s^{-1}sr^{-1}
\eqno{\rm B13}
$$

Finally, it is interesting to note that for reasonable confinement
regions, $L_{\rm e}\approx 10\,\rm kpc$, the ratio of the flux of
IC $\gamma$-rays from the CMB to the primary electron flux is
$$
{J_\gamma^{\rm IC}(>E)\over J_{\rm e}(>E)}
\approx 0.012
\left(E\over 1\,\rm TeV\right)^{1.15}
\eqno{\rm B14}
$$
which is less than unity for $E<40\,\rm TeV$. In fact this ratio is only
an upper limit because Klein-Nishina effects become important for IC on
the CMB at about $40\,\rm TeV$ (see above).

\bye